\documentclass[a4paper,fleqn,usenatbib]{mnras}

\usepackage{graphicx}	
\usepackage{amsmath}	
\usepackage{amssymb}	

\newcommand{\dw}{DE0630$-$18}
\newcommand{\aphotocentre}{$a_\mathrm{phot}$}

\title[Individual masses of \dw]{Individual Dynamical Masses of DENIS J063001.4$-$184014AB Reveal A Likely Young Brown Dwarf Triple\thanks{\textit{Based on observations made with ESO telescopes at the La Silla Paranal Observatory under programme IDs 086.C-0680, 088.C-0679, 090.C-0786, 092.C-0202.}}} 

\author[Sahlmann et al.]{
J.\ Sahlmann,$^{1,2,3}$\thanks{E-mail: jsahlmann@sciops.esa.int (JS)}
T.\ J.\ Dupuy,$^{4}$  
A.\ J.\ Burgasser,$^{5}$
J.\ C.\ Filippazzo,$^{3}$\newauthor
E.\ L.\ Mart\'{i}n,$^{6,7,8}$
D.\ C.\ Bardalez Gagliuffi,$^{9}$
C.\ Hsu,$^{5}$
P.\ F.\ Lazorenko,$^{10}$\newauthor
Michael C.\ Liu$^{11}$\\
$^{1}$RHEA Group for the European Space Agency (ESA), European Space Astronomy Centre (ESAC),\\ Camino Bajo del Castillo s/n, 28692 Villanueva de la Ca\~nada, Madrid, Spain\\
$^{2}$Independent researcher\\
$^{3}$Space Telescope Science Institute, 3700 San Martin Drive, Baltimore, MD 21218, USA\\
$^{4}$Institute for Astronomy, University of Edinburgh, Royal Observatory, Blackford Hill, Edinburgh, EH9 3HJ, UK\\
$^{5}$Center for Astrophysics and Space Science, University of California San Diego, La Jolla, CA, 92093, USA\\
$^{6}$Instituto de Astrof\'isica de Canarias (IAC), Calle V\'ia 
L\'actea s/n, E-38200 La Laguna, Tenerife, Spain\\
$^{7}$Departamento de Astrof\'isica, Universidad de La Laguna (ULL), 
E-38206 La Laguna, Tenerife, Spain\\
$^{8}$Consejo Superior de Investigaciones Cient\'ificas, E-28006 Madrid, Spain\\
$^{9}$Department of Astrophysics, American Museum of Natural History, Central Park West at 79th Street, NY 10024, USA\\
$^{10}$Main Astronomical Observatory, National Academy of Sciences of the Ukraine, Zabolotnogo 27, 03680 Kyiv, Ukraine\\
$^{11}$Institute for Astronomy, University of Hawaii, 2680 Woodlawn Drive, Honolulu, HI 96822, USA
}

\date{Accepted 2020 November 9. Received 2020 October 9; in original form 2020 August 5.}
\pubyear{2020}

\begin{document}
\label{firstpage}
\pagerange{\pageref{firstpage}--\pageref{lastpage}}
\maketitle

\begin{abstract}
The binary nature of the M8.5 dwarf \href{http://simbad.u-strasbg.fr/simbad/sim-basic?Ident=DENIS+J063001.4-184014&submit=SIMBAD+search}{DENIS J063001.4$-$184014AB} (\dw) was discovered with astrometric monitoring from the ground, which determined the unresolved photocentric orbit and the trigonometric parallax of the system. Here we present radial-velocity monitoring and resolved observations in the near-infrared with Keck aperture masking that allow us to measure the system's relative separation and brightness. By combining all available information, we determine the individual dynamical masses of the binary components to be $M_1 = 0.052^{+0.009}_{-0.008}$~$M_\mathrm{Sun}$ and $M_2 = 0.052^{+0.005}_{-0.004}$~$M_\mathrm{Sun}$, both firmly in the substellar regime. These masses are surprising given the object's M8.5 optical spectral type and equivalent absolute magnitude, and the significant difference in brightness between the components ($\Delta{K}$ = 1.74$\pm$0.06\,mag).  Our results suggest that \dw\ is a relatively young system ($\sim$200 Myr) with a secondary component that is itself a potentially unresolved binary. 
\end{abstract}

\begin{keywords}
brown dwarfs -- astrometry -- binaries: visual -- parallaxes -- stars: individual: \dw
\end{keywords}

\section{Introduction}
The masses of ultracool dwarfs, i.e.\ very low-mass stars and brown dwarfs, are difficult to measure precisely; yet they are necessary to refine our theoretical understanding of these objects and their physics \citep[e.g.][]{Zapatero-Osorio:2004fu, 2008ApJ...689..436L}. The monitoring of binary star motions that are governed by gravitational interaction give us the opportunity to determine the components' masses. Astrometric measurements of both the `absolute' positions in the sky and `relative' positions of the two components to each other give direct access to the individual masses, which can be further constrained with radial velocity monitoring. Such determinations have so far been made for a few dozen ultracool and brown dwarfs \citep[e.g.][]{Garcia:2017aa, 2017ApJS..231...15D} and recently for the directly-imaged giant extrasolar planet $\beta$~Pic~b \citep{Snellen:2018aa, 2019ApJ...871L...4D, 2020AJ....159...71N}.

DENIS J063001.4$-$184014 (hereafter \dw) was discovered as a late-type object by \citet{Phan-Bao:2008fr} and classified as an M8.5 dwarf from its optical spectrum.
It is part of a long-term astrometric monitoring campaign that makes use of the FORS2 optical camera mounted on the Very Large Telescope \citep{Sahlmann:2014aa}. The discovery and characterisation of the binary orbit with an orbital period of $3.067\pm0.006$\,years is described in \cite{Sahlmann:2015ac}. Its proximity ($\sim$20 pc) and expected orbital separation should make it possible to resolve the components with infrared (IR) adaptive optics (AO) facilities on 10m-class telescopes. We therefore initiated corresponding follow-up observations in 2014.

Here, we present a detailed characterisation and dynamical mass determinations for the components in the \dw\ binary, made possible by a newly-obtained relative position measurement complemented by radial-velocity monitoring.

\section{Observations and data reduction}
\subsection{Very Large Telescope/FORS2 astrometry}
 Here we do not present new FORS2 measurements, but we improve the accuracy of prior measurements by accounting for the updated camera calibrations presented in \citet{Lazorenko:2017aa} and \citet{2019A&A...629A.113L}. These improvements include the use of Gaia catalogues \citep{GaiaCollaboration:2016ab} for external astrometric calibration and accounting for the subtle changes in the camera's CCD chip locations.

{For the conversion of our differential astrometric measurements to ICRS we used Gaia DR2 and applied the transformation procedure
similar to that which we described in \citet{Lazorenko:2018aa}.
As the comparison epoch, we adopted  the average 
epoch $\bar T= 56249.955966$ MJD. 
Applying different polynomials for the transformation between
FORS2 and DR2 and different criteria for elimination of outliers we 
derived solutions with different samples of common stars
whose number varied between 21 to 57, just sufficient given
the number of coefficients between 10 and 28 per axis. 
We derived an RMS of $4$~mas for the
'FORS2-DR2' difference in positions, which exceeds its expected value
of about $1$~mas.  Such a large RMS value is probably due to
the different scale and relative rotation of the chips in the FORS2 detector,
which was not introduced in the transformation model whereas the difference
in zero points was taken into account as discussed in 
\citet{2019A&A...629A.113L}. }

{This transformation quality is however acceptable for the purpose of presenting the 
FORS2 positions in the ICRS after elimination of the
geometric field deformation, its rotation, and the calibration of the
pixel scale.} 

{Table \ref{tab:fors2_obs} contains the measured photocentre astrometry, which is given relative to the reference position
$\alpha_0= 97.507229262\degr$, $\delta_0= -18.672554677\degr$ 
at time  $\bar T$. 
We estimated that the actual accuracy of the reference position is 
 $\sim$5 mas, however the relative positions are about an order of magnitude more accurate.}

\begin{table*}
\caption{Individual FORS2 astrometric measurements of \dw\ for illustration. The complete table is available in electronic format. $E$ is the epoch number and $\Delta\alpha^{\star}_{m}$ and $\Delta\delta_{m}$ are the offsets in frame $m$ relative to the reference position $\alpha_0$, $\delta_0$ at time $\bar T$. The four $f$ coefficients are required to model DCR, as explained in Section \ref{sec:model}.}
\begin{tabular}{ccccccccccc}
\hline
\hline
$E$ & $m$ & $t_m$ & $\Delta\alpha^{\star}_{m}$ & $\sigma_{\alpha^{\star}_{m}}$ & $\Delta\delta_{m}$ & $\sigma_{\delta_{m}}$ & $f_{1,x,m}$ & $f_{2,x,m}$ & $f_{1,y,m}$ & $f_{2,y,m}$ \\
 & & (MJD) & (mas) & (mas) & (mas) & (mas) & & & & \\
\hline
1 & 1 & 55537.27016 & -496.105 & 0.738 & 841.776 & 0.825 & 0.08339 & 0.10735 & 0.10616 & 0.13666 \\
1 & 2 & 55537.27077 & -496.548 & 0.915 & 841.982 & 1.020 & 0.08701 & 0.11004 & 0.10635 & 0.13450 \\
1 & 3 & 55537.27138 & -496.512 & 0.865 & 842.772 & 0.960 & 0.09060 & 0.11259 & 0.10652 & 0.13237 \\
1 & 4 & 55537.27200 & -496.613 & 0.769 & 842.034 & 0.850 & 0.09418 & 0.11501 & 0.10667 & 0.13027 \\
1 & 5 & 55537.27261 & -495.803 & 0.749 & 842.397 & 0.847 & 0.09774 & 0.11732 & 0.10681 & 0.12820 \\
\hline
\end{tabular}
\label{tab:fors2_obs}
\end{table*}

\subsection{Keck/NIRC2 LGS AO}
We first observed \dw\ with the laser guide star adaptive optics (LGS AO) system at the Keck~II telescope \citep{2004SPIE.5490..321B, 2006PASP..118..297W, 2006PASP..118..310V} on 2014~Mar~14~UT.  We obtained data using the 9-hole non-redundant aperture mask installed in the filter wheel of NIRC2 \citep{2006SPIE.6272E.103T} and analyzed these data using the same pipeline as in our previous work \citep[e.g.][]{Dupuy:2009lk, 2015ApJ...805...56D, 2017ApJS..231...15D}. Interferograms taken in both the $H$ and $K$ bands, from the standard Mauna Kea Observatories (MKO) filters \citep{2002PASP..114..169S, 2002PASP..114..180T}, showed a significant detection of a binary. Fitting the closure phases derived from the higher-quality $K$-band data gave a separation of $48\pm3$\,mas, position angle (PA) of $302\degr\pm3\degr$ and flux ratio of $1.88\pm0.13$\,mag, where errors were computed using a Monte Carlo method that accounted for the measured closure phase errors. We observed \dw\ again on 2017~Mar~20~UT, this time in the MKO $K_S$ band with a PSF calibrator observed immediately after the science target. This allowed us to measure more accurate binary parameters with a separation of $36.6\pm1.3$\,mas, PA of $291\fdg5\pm1\fdg2$, and flux ratio of $1.74\pm0.06$\,mag (Table \ref{tab:nrm_obs}). 

The PA measured from the 2014 data {(Table \ref{tab:nrm_obs2})} is highly inconsistent ($\approx$10\degr\ off, {see Figure \ref{fig:rel_orbit}}) compared to the prediction from the absolute astrometric orbit from FORS data. Such a discrepancy is unprecedented in our experience with other data obtained for similar binaries, even accounting for the lack of a PSF calibrator in 2014. Ultimately, we choose to exclude the 2014 measurements from our analysis, and the source of the PA discrepancy remains unknown.

\begin{table}
\caption{Results of Keck aperture masking (10 March 2017).}
\centering
\begin{tabular}{c c  r r}
\hline
\hline
Sep. & (mas) & $36.6\pm1.3$ \\
PA & (\degr) &  $291.5\pm1.2$\\
$\Delta K_S$ & (mag)  &$1.74 \pm 0.06 $\\
\hline
\end{tabular}
\label{tab:nrm_obs}
\end{table}

\begin{table}
\caption{{Results of Keck aperture masking (13 March 2014). These data were not used in the analysis.}}
\centering
\begin{tabular}{c c r r r}
\hline
 & & $H$ & $K_s$\\
 \hline
Sep. & (mas) & $48.1\pm2.7$ & $51.8\pm2.2$\\
PA & (\degr) & $302.0\pm2.4$ &  $304.3\pm3.2$ \\
$\Delta$Mag & (mag) & N/A &$1.88 \pm 0.13 $ \\
$\Delta$Mag & (mag) & $1.87 \pm 0.19$ &N/A  \\
\hline
\end{tabular}
\label{tab:nrm_obs2}
\end{table}

\subsection{High-resolution infrared spectroscopy}
\dw\ was observed with the Keck II Near InfraRed Spectrometer (NIRSPEC; \citealt{2000SPIE.4008.1048M}) on four nights: 2016 November 16, 2017 February 6, 2017 March 22 and 2017 December 7 (UT). For each observation we used the N7 order-sorting filter and 0$\farcs$432-wide slit to obtain 2.00--2.39~$\micron$ spectra over orders 32--38 with $\lambda/\Delta\lambda$= 20,000 ($\Delta{v}$ = 15~km/s) and dispersion of 0.315~{\AA}~pixel$^{-1}$. Two dithered exposures of 1000~s (2016 Nov) or 600~s (2017 Feb, Mar and Dec) each were obtained, along with observations of the nearby A0~V star HD~49529 ($V$ = 8.09).  Flat field lamp, arc lamp and dark frame exposures were obtained at the start of each night for calibration.  
The NIRSPEC data were reduced and forward-modeled as described in \citet{Burgasser:2015aa}, \citet{2020NatAs.tmp...43T} and Hsu et al.\ (in prep.), 
using a modified version of NIRSPEC Data Reduction Pipeline for reduction \citep{2016SPIE.9910E..2ET}, telluric absorption models from \citet{2014A&A...568A...9M}, and the BT-Settl solar-metallicity atmosphere models \citep{Allard:2011aa} for the target spectrum. The analysis of the NIRSPEC data focused on the order 33, which covers both the CO $\nu$ = 2-0 band at 2.29~$\mu$m and telluric CO and H$_2$O absorption used to refine the wavelength solution. Our forward-modeling method first fits a associated A0~V star spectrum to measure the instrumental line-spread function (LSF) modeled as a Gaussian broadening kernel, airmass, and precipitable water vapor (pwv). We then fit the stellar spectrum to a nine parameter model using a Markov Chain Monte Carlo (MCMC) algorithm, including four stellar parameters (effective temperature T$_{eff}$, surface gravity $\log{g}$, projected rotational velocity $v\sin{i}$, and radial velocity RV), two meteorological parameters (airmass and pwv), and three nuisance parameters (flux and wavelength offsets, and a noise scale factor). The rotational broadening profile assumes a limb-darkening coefficient of 0.6 \citep{1992oasp.book.....G}. We also fit out a 10th-order polynomial continuum correction at the end of each MCMC step. The MCMC was run with 50 walkers and 2,000 steps, with a sigma-clipping mask threshold of 2.5 $\sigma$ used to reject outlying pixels beyond the 1,000 step (less than 2\% of the pixels were removed). Final fit parameters were determined from the mean and distribution of values in the final 800 steps, and a barycentric correction was applied to the inferred RV. 
We also re-evaluated the UVES data from \citet{Sahlmann:2015ab} following similar methods, focusing on the 819~nm Na~I doublet, which falls in a relatively high S/N = 34 region of the observed data.

\begin{table*}
\caption{Radial and Rotational Velocities from UVES \& NIRSPEC Observations.}
\centering
\begin{tabular}{c c c r r r r r}
\hline
\hline

Instrument & MJD & Median & T$_{eff}$ & log\,g & RV & $v \sin i$ \\
 && S/N & (K) & (cm/s$^2$) & (km/s) & (km/s) \\
 \hline
VLT/UVES & 56568.36294 & 34 & 2634$\pm$44 & 4.54$\pm$0.11 & $-$10.13$\pm$0.71 & 9.3$\pm$1.1 \\
Keck/NIRSPEC & 57708.49016  & 33 & 2735$\pm$7 & 5.49$\pm$0.02 & $-$9.65$\pm$0.27 & 11.2$\pm$0.7 \\
Keck/NIRSPEC & 57790.27401 &  68 & 2758$\pm$7 & 5.49$\pm$0.01 &$-$7.48$\pm$0.17 & 10.4$\pm$0.7 \\
Keck/NIRSPEC & 57834.22386 &  50 & 2750$\pm$5 & 5.49$\pm$0.01 & $-$5.90$\pm$0.14 & 10.9$\pm$0.8  \\
Keck/NIRSPEC & 58094.49858 & 52 & 2752$\pm$5 & 5.49$\pm$0.01 & $-$7.82$\pm$0.15 & 10.2$\pm$0.6 \\
\hline
\end{tabular}
\label{tab:nirspec}
\end{table*}

\begin{figure}
\centering
\includegraphics[width=\linewidth]{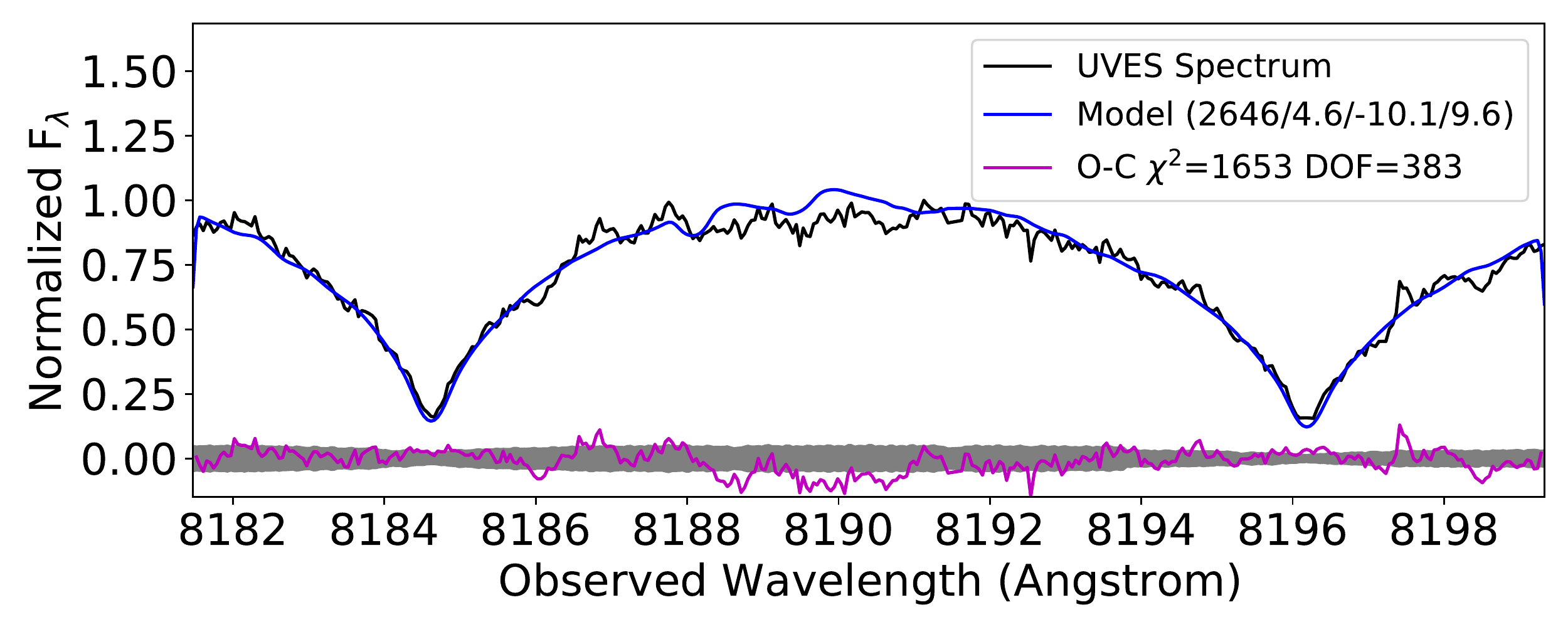}
\includegraphics[width=\linewidth]{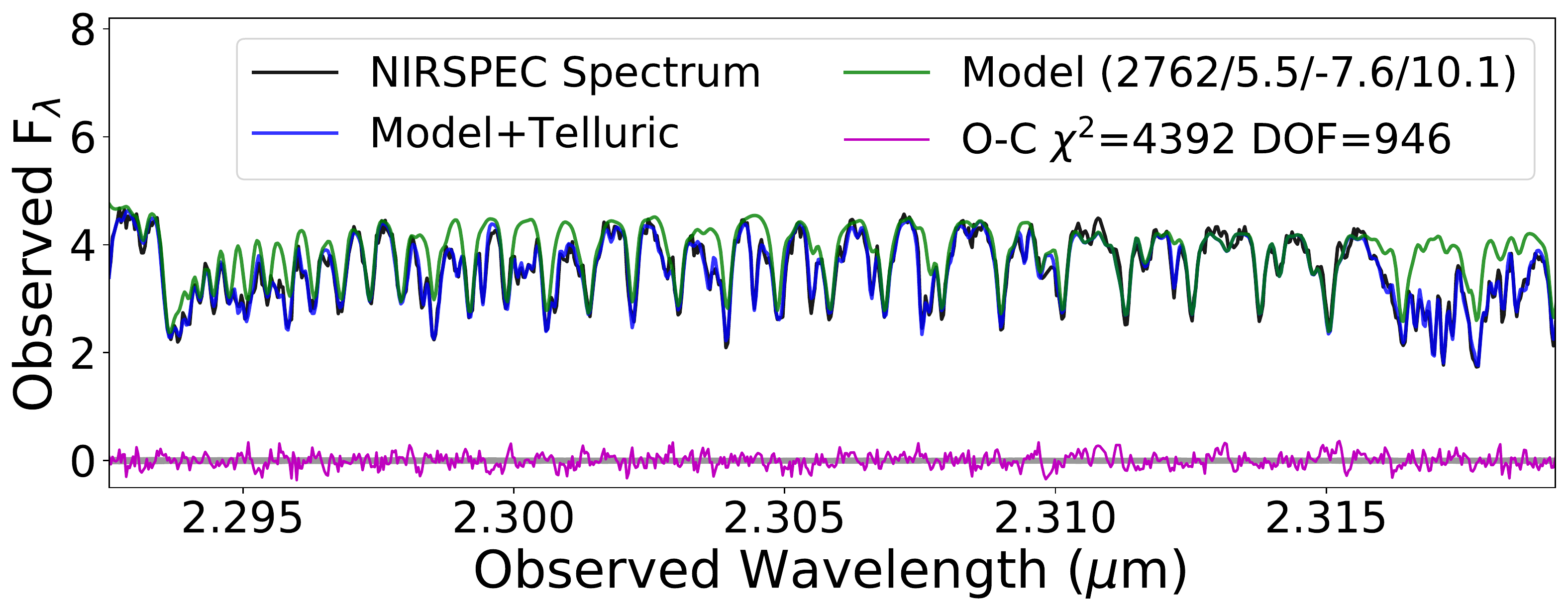}
\caption{Forward model fits to the UVES spectrum (top) and NIRSPEC spectrum (bottom) of DE0630-18. Both panels show the data in black, the best-fit model in blue, and the difference spectrum in magenta (overlapping the 1$\sigma$ uncertainty range in grey. The dashed lines in the UVES plot indicate masked-out telluric absorption features, while the green line in the NIRSPEC plot shows the model without telluric absorption included.}
\label{fig:highres}
\end{figure}

Table~\ref{tab:nirspec} summarizes the resulting atmospheric parameters and radial and rotational velocities determined by this analysis, while Figure~\ref{fig:highres} displays fits for one epoch each of the UVES and NIRSPEC data.
Both the effective temperature ($T_{eff}$ = 2634$\pm$44~K for UVES, ${\langle}T_{eff}{\rangle}$ = 2750$\pm$13~K for NIRSPEC) and rotational velocity measurements ($v\sin{i}$ = 9.3$\pm$1.1~km/s for UVES, ${\langle}v\sin{i}{\rangle}$ = 10.6$\pm$0.7~km/s for NIRSPEC) of \dw\ are consistent across all epochs. The surface gravity shows a full dex discrepancy between the UVES ($\log{g}$ = 4.54$\pm$0.11) and NIRSPEC ($\langle\log{g}\rangle$ = 5.49$\pm$0.01) data, an issue previously noted in optical and infrared spectral modeling of the young eclipsing brown dwarf binary SPEC~J1510$-$2828AC \citep{2020NatAs.tmp...43T}. The radial velocity measurements, which also vary significantly between epochs, is fortunately insensitive to the atmospheric parameters. We obtain statistically identical results if we constrain the fits to either the low or high values of $\log~g$ inferred. 

\section{Data analysis}  

\subsection{Parallax correction}
In the pre-Gaia work of \citet{Sahlmann:2014aa} we determined the correction from relative to absolute parallax on the basis of a Galaxy model. Here, we can improve this by deriving 
the offset  $\Delta \varpi$ for our relative parallaxes $\varpi_F$ using the absolute Gaia DR2 parallaxes $\varpi_G$ \citep{Gaia-Collaboration:2018ae}. 

As described in \citet{Lazorenko:2014aa}, the astrometric reduction was made with different values of the model parameter $k$, which determines the degree of the bivariate polynomial used to map individual frames to each other, and corresponding radii $R$ of the reference field
(up to $\sim$2\arcmin\ for $k=16$).
The final solution is taken as the weighted average of these individual solutions.
Accordingly, we derived the parallax correction $\Delta \varpi$ using varying values of $k$ and $R$ by cross-matching the FORS2 stars and Gaia DR2 sources.

Many of the FORS2 reference stars lie at or beyond the faint end of Gaia DR2, thus were not used. For $k=10$ and $k=14$ we matched 32 and 63 reference stars, respectively, and derived $\Delta \varpi$ as the simple arithmetic average of differences $\varpi_G -\varpi_F$. After rejecting a few Gaia sources with abnormally large parallax uncertainties, we obtained corrections of $\Delta \varpi = 0.355 \pm 0.076$~mas and $\Delta \varpi = 0.331 \pm 0.059$~mas, respectively. These estimates are in good agreement and we adopted their average value of $\Delta \varpi = 0.343 \pm 0.068$~mas as the final estimate, which is also compatible with the initial estimate of $0.42 \pm 0.04$~mas in \citet{Sahlmann:2014aa}.

\subsection{Converting MKO $K$-band magnitude difference to the FORS2 $I$-band}\label{sec:delta_mag}
To convert the photocentric FORS2 astrometry into barycentric astrometry a measurement of the magnitude difference in the FORS2 filter ($\Delta I_\mathrm{FORS2}$) is required. 
The Keck aperture mask observations yield a measurement of $\Delta K$, which we needed to convert to $\Delta I_\mathrm{FORS2}$.
We used the \url{https://github.com/BDNYC/BDNYCdb} database of ultracool spectra and the spectral energy distribution (SED) tool available at \url{https://github.com/hover2pi/sedkit} \citep{Filippazzo:2015aa} to generate synthetic magnitudes in the $I_\mathrm{FORS2}$ and $K_S$
bands for a library of 141 M and L dwarfs, which includes 75 field objects and 66 low-gravity objects. For every source synthetic magnitudes were computed from its SED. 

\begin{figure}
\centering
\includegraphics[width=\linewidth]{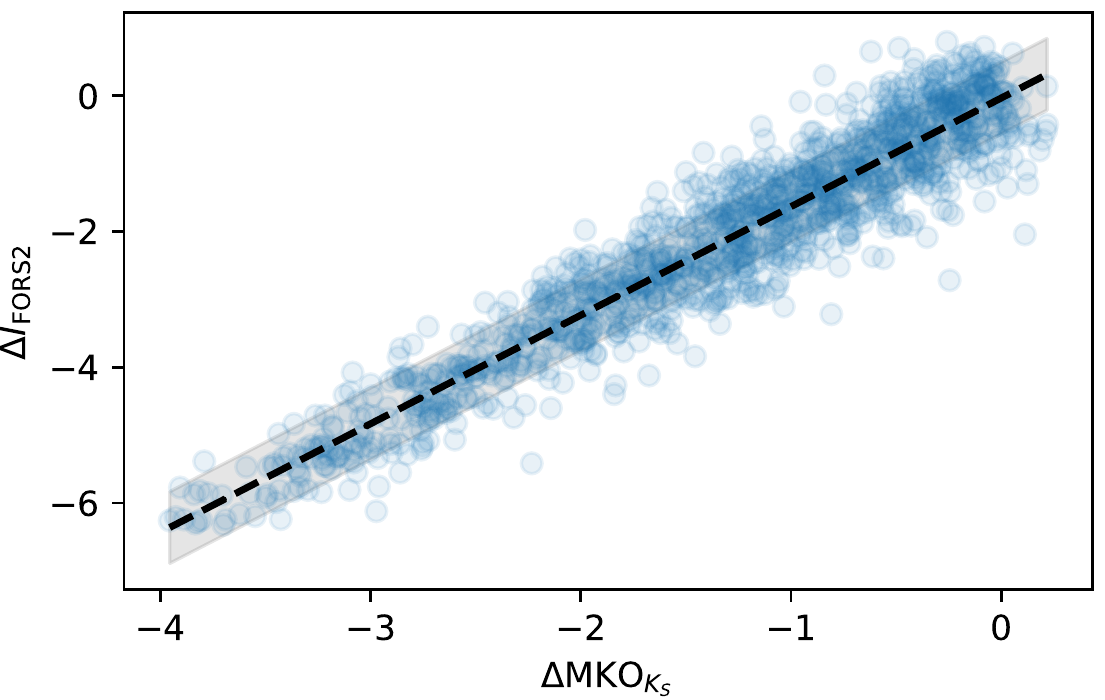}
\caption{Synthesised magnitude differences in $I_\mathrm{FORS2}$ and $\mathrm{MKO}_{K_S}$ and the corresponding linear fit (dashed line).}
\label{fig:mag_conv_1}
\end{figure}

The 1230 pairwise combinations of sources with spectral types between M6 and L8 were used to estimate the magnitude differences shown in Figure \ref{fig:mag_conv_1}. The relationship is well approximated by a straight line and we used the coefficients of the linear fit (not accounting for individual data point uncertainties) to convert the measured $\Delta K_S$
to  $\Delta I_\mathrm{FORS2}$, and we added the residual RMS of the fit (0.53 mag) in quadrature to the uncertainty in $\Delta K_S$.
The measured value of  $\Delta K_S = 1.74 \pm 0.06$\,mag (Table \ref{tab:nrm_obs}) is then converted to $\Delta I_\mathrm{FORS2}= 2.82 \pm 0.53$.

\subsection{Combined astrometric model}\label{sec:model}
The astrometric measurements of the target are $\alpha^{\star}_m= \alpha_m \cos{\delta}$ and $\delta_m$, corresponding to Right Ascension and Declination, respectively, in frame $m$ at time $t_m$ relative to the reference frame of background stars. These are modeled with seven free parameters $\Delta\alpha^{\star}_0, \Delta\delta_0, \mu_{\alpha^\star}, \mu_\delta, \varpi$, $d$, and $\rho$ as:
\begin{equation}\label{eq:axmodel}
\begin{array}{l@{\hspace{0mm}}l@{\hspace{0mm}}l@{\hspace{0mm}}l}
\!\alpha^{\star}_{m} =& \Delta \alpha^{\star}_0 + \mu_{\alpha^\star} \, t_m + \varpi \, \Pi_{\alpha,m} &-\;\; \rho\, f_{1,x,m}&-\;\; d\, f_{2,x,m} \\
\delta_{m} = &{\Delta \delta_0 + \,\mu_\delta      \,  \;                      t_m \;+ \varpi \, \Pi_{\delta,m}}  &{+\;\; \rho \,f_{1,y,m}}&{+\;\; d \,f_{2,y,m}},
\end{array}
\end{equation}
where $\Delta\alpha^{\star}_0, \Delta\delta_0$ are the coordinate offsets, $\mu_{\alpha^\star}, \mu_\delta$ are the proper motions, and the parallactic motion is expressed as the product of relative parallax $\varpi$ and the parallax factors $\Pi_\alpha, \Pi_\delta$. The atmospheric differential chromatic refraction (DCR) is modelled with the free parameters $\rho$ and $d$ \citep{Lazorenko:2011lr, Sahlmann:2014aa} and the 
coefficients $f_{1}$ and $f_{2}$, where {the latter are fully determined as} a function of zenith angle, temperature, and pressure \citep{Lazorenko:2006qf, Sahlmann:2013ab, Sahlmann:2016aa}. {The DCR treatment does not involve the estimation of source colours, instead $\rho$ and $d$ are empirical free model parameters that correspond to the effective colour of the target relative to the average reference star.}

The Keplerian orbit model adds an additional seven free parameters to the model.
These are the eccentricity $e$, the argument of periastron $\omega$, the orbital period $P$, the longitude of ascending node $\Omega$, the orbital inclination $i$, the time of periastron passage $T_\mathrm{P}$, and the semi-major axis of the photocentre orbit \aphotocentre.
Since we estimated the  
magnitude difference between the components in the filter bandpass in Section \ref{sec:delta_mag}, we can relate the photocentre orbit size to the barycentre orbit size $a_1$ of the primary, similarly to \citet{Sahlmann:2020}. 
In the photocentre astrometry model that is applied to the FORS2 data we also include the astrometric nuisance  parameters $s_\alpha$ and $s_\delta$ which can account for excess noise in the astrometry \citep{Sahlmann:2013ab}.

The five available radial velocity measurements of the primary (Table \ref{tab:nirspec}) were modelled in a standard way. As discussed in \citet{Sahlmann:2015ab} we do not expect significant offsets between the UVES and NIRSPEC instruments and can therefore model those data jointly. Since the radial velocity variation of the primary is fully characterised by the orbital parameters above, the inclusion of radial velocities adds only the systemic velocity $\gamma$ as free parameter. 

Finally, we included the relative astrometric measurement from the resolved Keck aperture mask observation (Table \ref{tab:nrm_obs}). This allows us to determine model-independent masses because we can now directly adjust the primary and companion mass as free parameters ($M_1$ and $M_2$) instead of \aphotocentre\ or $a_1$. Apart from that, this step does not introduce any new parameters because the relative orbit is fully determined by the orbital parameters and the component masses.

\subsection{MCMC analysis}
We used the astrometric and orbital parameters from \citet{Sahlmann:2015ac} and reasonable guesses for the component masses and systemic radial velocity 
as starting values for a Markov Chain Monte Carlo (MCMC) analysis similar to that described in \citet{Sahlmann:2020}.
We used the \texttt{emcee} package \citep{Foreman-Mackey:2013aa} to implement the MCMC and expressed the binary model using  \texttt{pystrometry} \citep{johannes_sahlmann_2019_3515526}\footnote{\url{https://github.com/Johannes-Sahlmann/pystrometry}} with the parameter vector $\theta$ composed of $\Delta \alpha^\star_0$, $\Delta \delta_0 $, $\varpi$, $\mu_{\alpha^\star}$, $\mu_\delta$, $\rho$, $d$,  $P$, $e$, $\omega$,
$T_P$,
$s_\alpha$, $s_\delta$, 
$M_\mathrm{1}$, $M_\mathrm{2}$, $i$
and $\Omega$, $\gamma$.

The magnitude difference $\Delta I_\mathrm{FORS2}$ and the parallax correction $\Delta \varpi$ are incorporated as Gaussian priors in the MCMC \citep[see][]{Sahlmann:2020}. 
Finally, the reference time $T_\mathrm{Ref}$ and the absolute coordinates enter the model as constants. 
The full model has 18 free parameters and two additional parameters constrained by priors.

\subsection{Results}
\begin{figure}
\centering
\includegraphics[width=\linewidth]{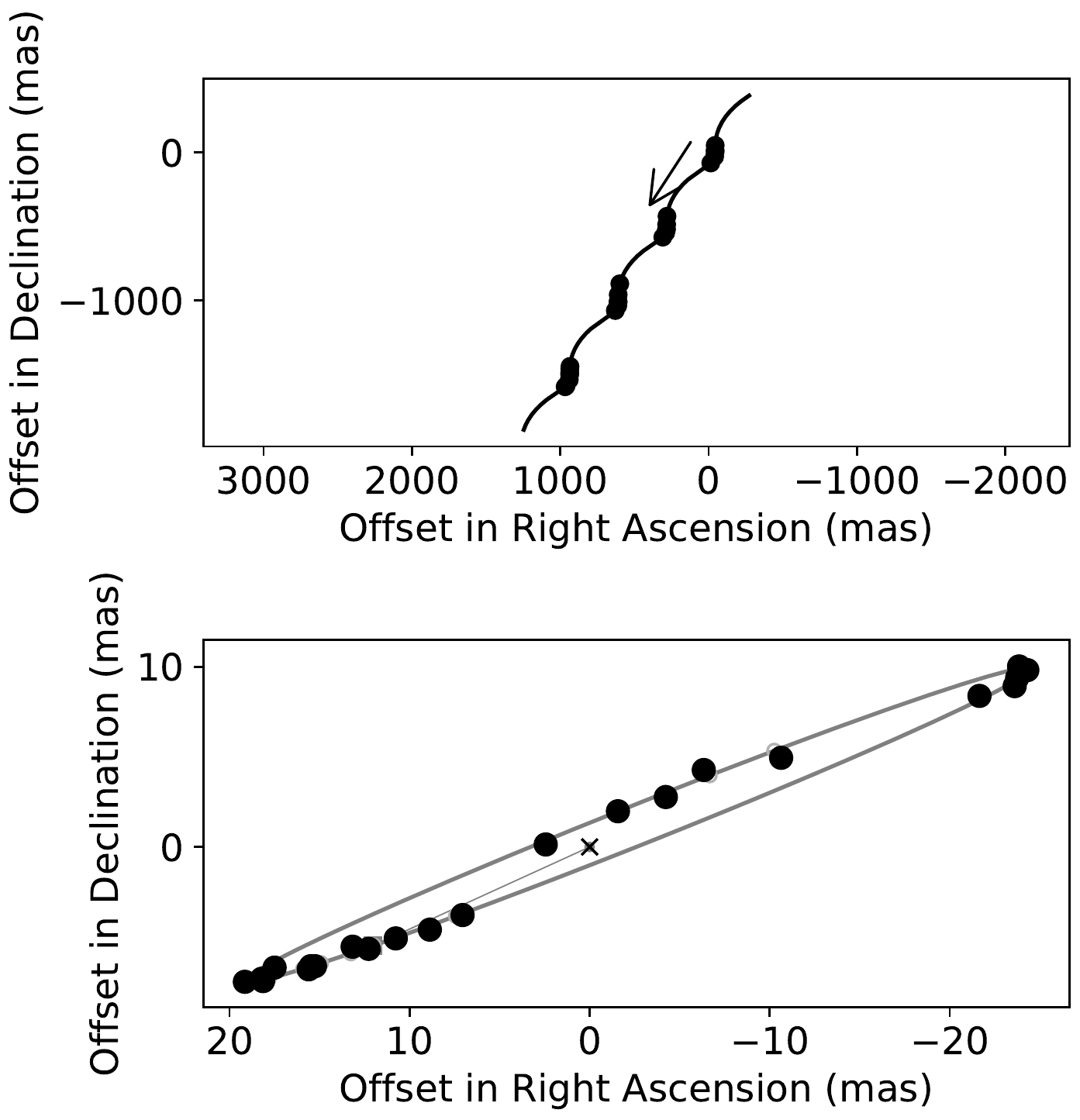}
\caption{Best fit model of FORS2 parallax/proper motion (top) and photocentric orbit (bottom). {The black circles indicate the epoch-averaged FORS2 measurements with uncertainties that are smaller than the symbol size.} }
\label{fig:orbit}
\end{figure}

\begin{figure}
\centering
\includegraphics[width=\linewidth]{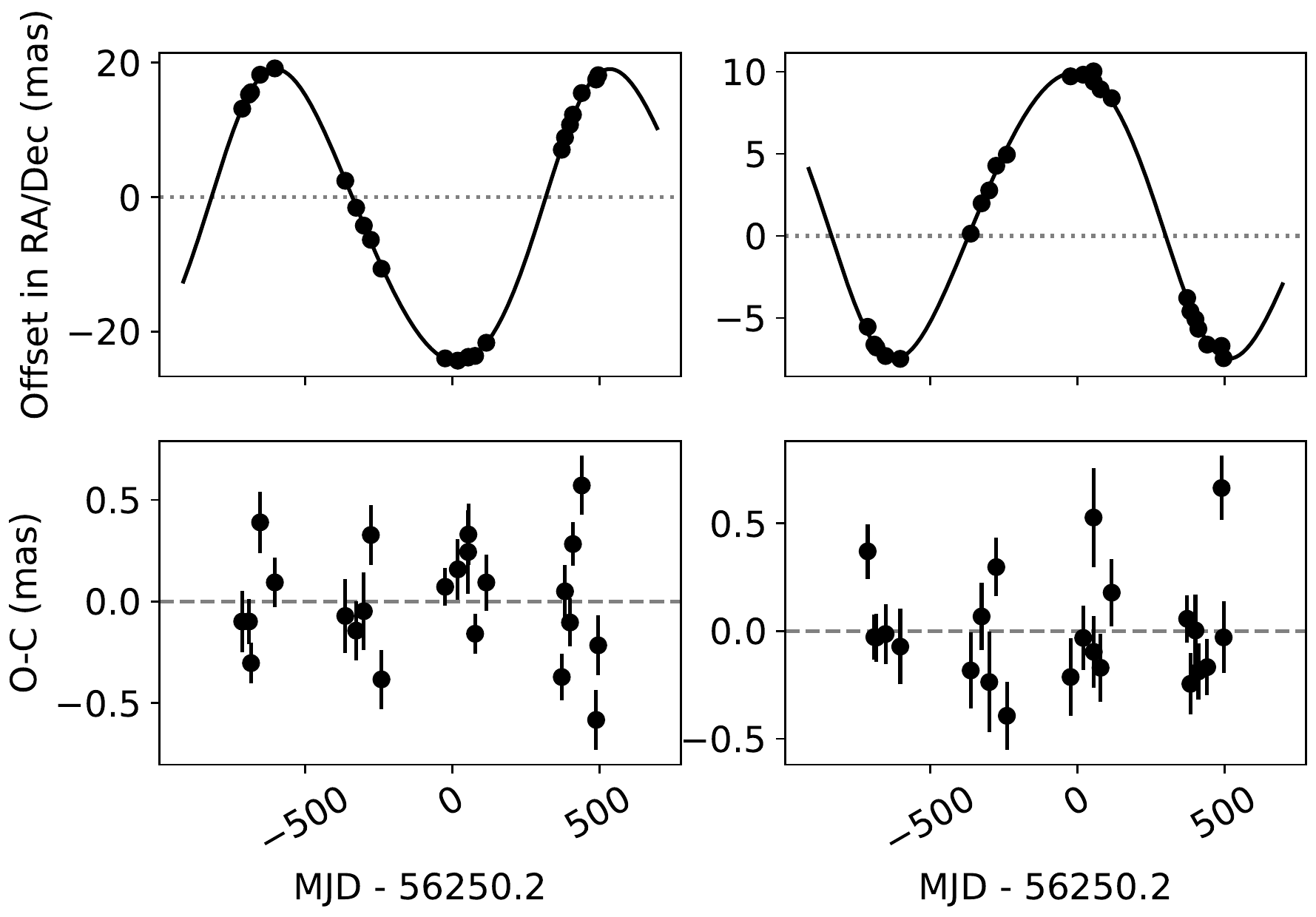}
\caption{Top: Best fit model of FORS2 photocentric orbit as a function of time. Bottom: O-C residuals. The fit quality (0.27 mas RMS) is marginally deteriorated compared to the astrometry-only fit of \citet[][Fig.\ 4, 0.24 mas RMS]{Sahlmann:2015ac}.}
\label{fig:orbit_time}
\end{figure}

\begin{figure}
\centering
\includegraphics[width=\linewidth]{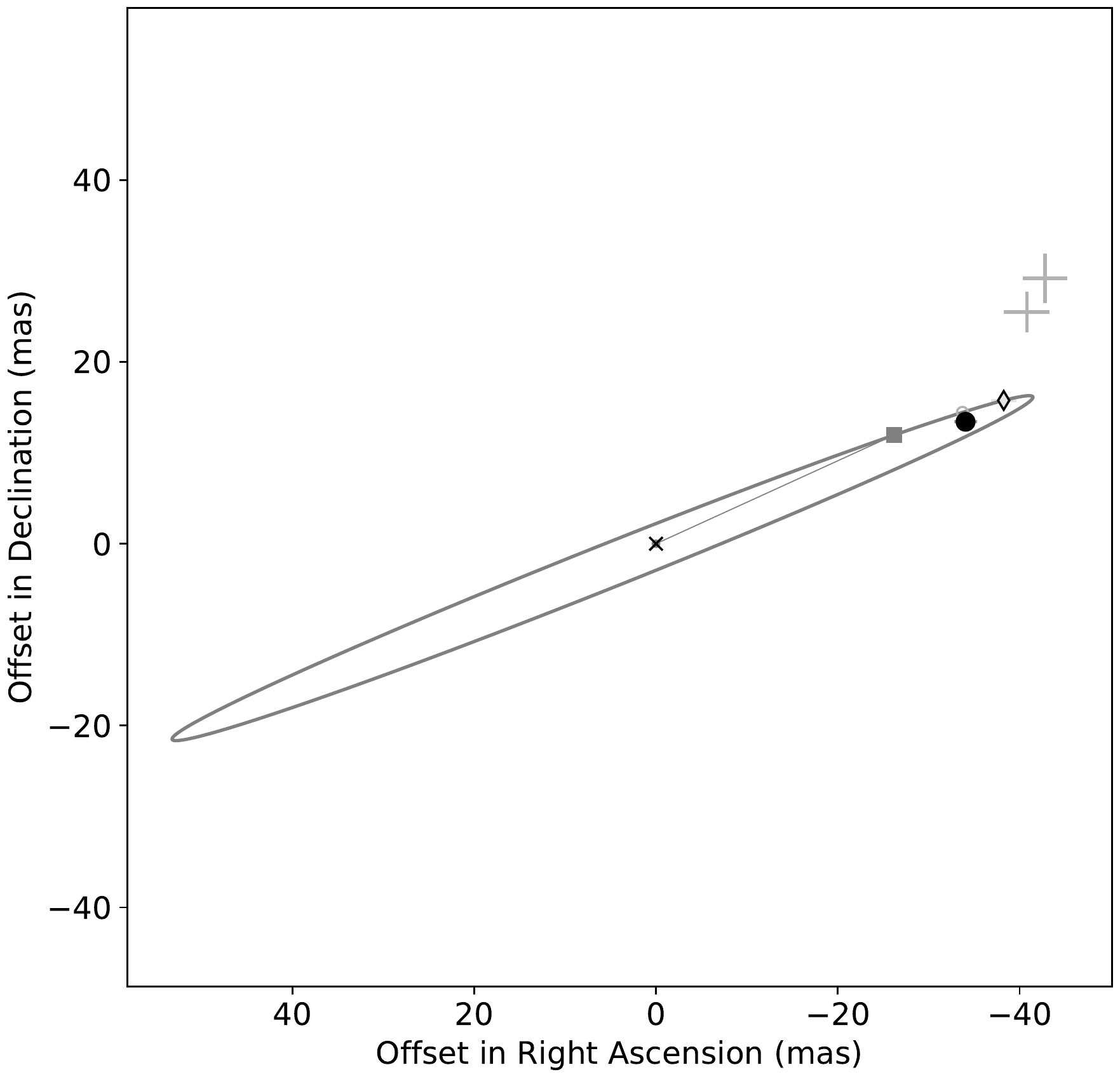}
\caption{Best fit model of the relative orbit. The black symbol shows the single Keck aperture masking measurement. {The grey crosses mark the discarded 2014 measurements and the diamond marks their expected position.} }
\label{fig:rel_orbit}
\end{figure}

\begin{figure}
\centering
\includegraphics[width=\linewidth]{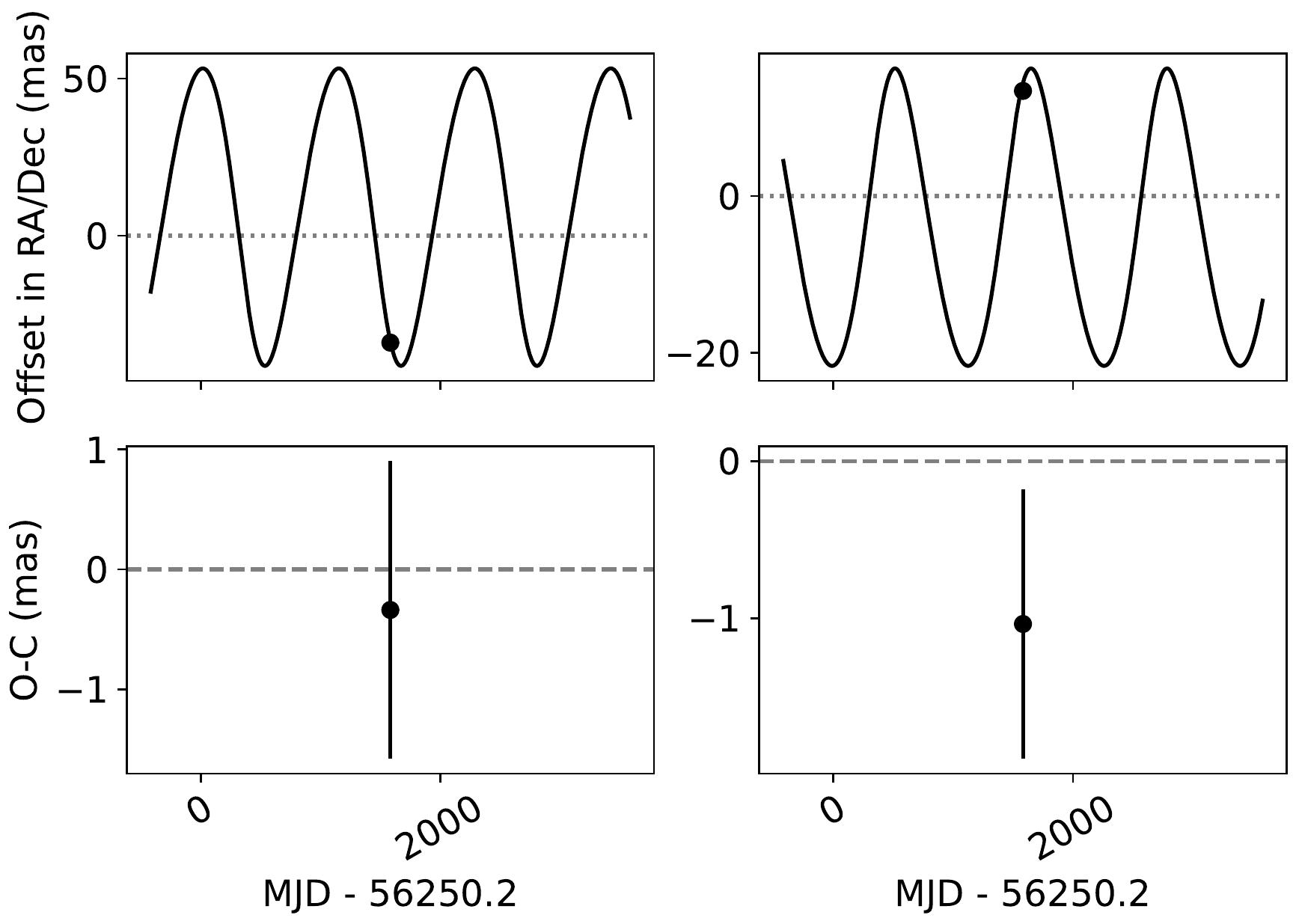}
\caption{Top: Best fit model of relative orbit as a function of time. Bottom: O-C residuals. The black symbol shows the single used Keck aperture masking measurement. }
\label{fig:orbit_time_relative}
\end{figure}

\begin{figure}
\centering
\includegraphics[width=\linewidth]{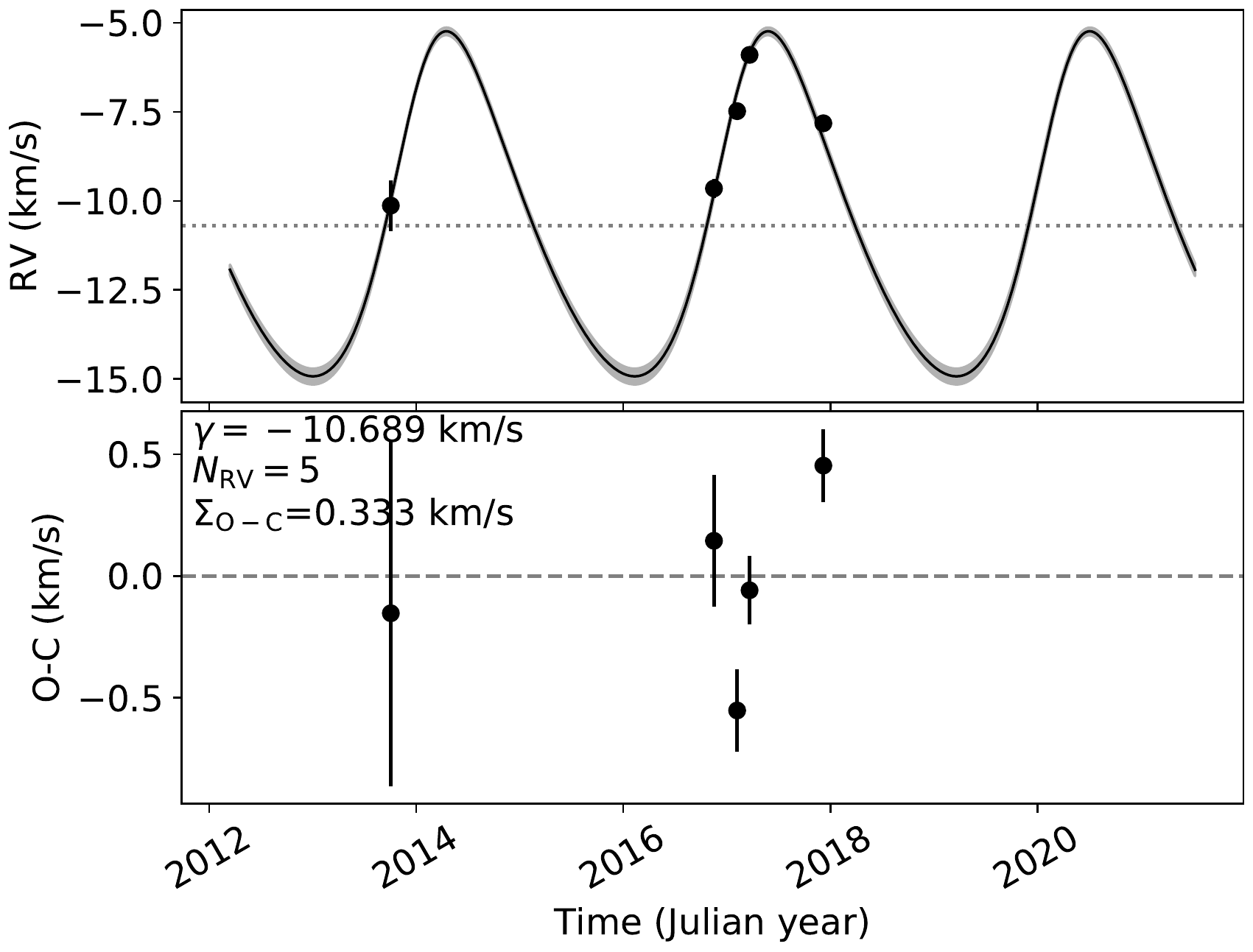}
\caption{Top panel: RV measurements (black symbols; the UVES data are from $\sim$2014 and the remaining data are from NIRSPEC) and the best-fit orbit (solid curve). The grey area corresponds to the 1-sigma equivalent range of RV values filled by random draws from the posteriors. The systemic velocity is shown by the horizontal dashed line. Bottom panel: O-C residuals.}
\label{fig:rv}
\end{figure}

In our analyses we are always using all the individual frame data for the model fitting. For better visualisation of the results, however, we display only the epoch averages in the figures.
Figure \ref{fig:orbit} shows the photocentric orbit of the binary in the sky and Figure \ref{fig:orbit_time} shows the same as a function of time. Figures \ref{fig:rel_orbit} and \ref{fig:orbit_time_relative} show the relative orbit with the Keck aperture mask measurement and Figure \ref{fig:rv} shows the radial velocity curve from the UVES and NIRSPEC measurements. 

Table \ref{tab:solution} lists the adopted solution parameters determined as the median of the posterior distributions with 1 $\sigma$-equivalent confidence intervals. The results are unexpected in the sense that the primary mass of $0.052^{+0.009}_{-0.008}$\,$M_\mathrm{Sun}$ is very low for an M8.5 dwarf and that the companion has essentially the same mass as its host. 
For comparison, the mean of seven dynamical masses for M8--M8.5 dwarfs measured by \citet{2017ApJS..231...15D} is 0.090\,$M_\mathrm{Sun}$, within $1 \sigma$ of the \emph{total} mass of the \dw\ system $0.104^{+0.013}_{-0.012}$,$M_\mathrm{Sun}$.

To estimate the dependency of this result on the empirically constrained $\Delta$mag parameter, we repeated the same analysis with larger magnitude differences, i.e.\ $\Delta I_\mathrm{FORS2}'=\Delta I_\mathrm{FORS2}+1.5$ mag and $\Delta I_\mathrm{FORS2}''=\Delta I_\mathrm{FORS2}+10$ mag. The results are reported in Tables \ref{tab:solution2} and \ref{tab:solution3}, respectively, and we concluded that our main results do not depend on the exact value of the magnitude difference, because the derived masses agree within the uncertainties.

The overall fit quality (photocentre motion, relative separation, RV) is comparable and cannot be used to identify the more likely scenario among the following cases:

\begin{itemize}
\item Nominal $\Delta I_\mathrm{FORS2}$:
This case corresponds to the `nominal' optical flux ratio derived from the measured $K$-band flux ratio and the empirical relationship from Section \ref{sec:delta_mag}.
The primary mass and the companion mass are identical to each other within the errors. 
The observational data appear not to be fully compatible with the prior on $\Delta I_\mathrm{FORS2}$ (=$2.8 \pm 0.5$ mag) since the corresponding posterior is shifted towards a higher value (the median deviates by +0.7 mag).

\item $\Delta I_\mathrm{FORS2}'=\Delta I_\mathrm{FORS2}+1.5$ mag:
The primary is marginally more massive than the companion (Table \ref{tab:solution2}).
The prior on $\Delta I_\mathrm{FORS2}'$ (=$4.3 \pm 0.5$ mag) is better fulfilled and the posterior's median deviates only by +0.3 mag.

\item $\Delta I_\mathrm{FORS2}''=\Delta I_\mathrm{FORS2}+10$ mag:
In this case the companion is essentially dark and photocentre and barycentre motion coincide.
Table \ref{tab:solution3} tabulates the median posterior values.
The secondary is slightly less massive than in the previous case and the primary mass remains unchanged.
The prior on $\Delta I_\mathrm{FORS2}''$ (=$12.8 \pm 0.5$) is fulfilled.
\end{itemize}

\begin{table}
\caption{Solution derived from the MCMC.}
\centering
\begin{tabular}{cr}
\hline\hline
Parameter & Value \\
\hline
$\Delta \alpha^\star_0$ (mas) & $574.59^{+0.17}_{-0.17}$ \\[3pt]
$\Delta \delta_0 $ (mas) & $-902.60^{+0.08}_{-0.08}$ \\[3pt]
$\varpi_\mathrm{abs}$ (mas) & $51.34^{+0.09}_{-0.09}$ \\[3pt]
$\mu_{\alpha^\star}$ (mas yr$^{-1}$) & $326.74^{+0.05}_{-0.05}$ \\[3pt]
$\mu_\delta$ (mas yr$^{-1}$) & $-502.65^{+0.03}_{-0.03}$ \\[3pt]
$\rho$ (mas) & $19.91^{+0.71}_{-0.72}$ \\[3pt]
$d$ (mas) & $-26.21^{+0.60}_{-0.60}$ \\[3pt]
$P$ (day) & $1135.14^{+1.67}_{-1.74}$ \\[3pt]
$P$ (yr) & $3.108^{+0.005}_{-0.005}$ \\[3pt]
$\Omega$ ($^\circ$) & $111.65^{+0.12}_{-0.12}$ \\[3pt]
$\gamma$ (m s$^{-1}$) & $-10689.43^{+119.15}_{-123.47}$ \\[3pt]
$\lambda_\mathrm{ref}$ ($^\circ$) & $-176.94^{+0.21}_{-0.20}$ \\[3pt]
$\sqrt{e}\sin\omega$ () & $-0.31^{+0.01}_{-0.01}$ \\[3pt]
$\sqrt{e}\cos\omega$ () & $0.29^{+0.01}_{-0.01}$ \\[3pt]
$\sqrt{M_2}\sin{i}$ ($M_\mathrm{Jup}$) & $7.35^{+0.32}_{-0.30}$ \\[3pt]
$\sqrt{M_2}\cos{i}$ ($M_\mathrm{Jup}$) & $-0.35^{+0.02}_{-0.02}$ \\[3pt]
$s_\alpha$ (mas) & $0.22^{+0.08}_{-0.10}$ \\[3pt]
$s_\delta$ (mas) & $0.16^{+0.10}_{-0.11}$ \\[3pt]
$e$ () & $0.183^{+0.007}_{-0.008}$ \\[3pt]
$\omega$ ($^\circ$) & $-46.76^{+1.80}_{-1.72}$ \\[3pt]
$i$ ($^\circ$) & $92.71^{+0.15}_{-0.15}$ \\[3pt]
$T_P$ (day) & $56660.70^{+5.49}_{-5.17}$ \\[3pt]
\aphotocentre\ (mas) & $23.55^{+0.06}_{-0.06}$ \\[3pt]
$a_1$ (mas) & $25.46^{+0.80}_{-0.61}$ \\[3pt]
$a_\mathrm{rel}$ (mas) & $51.42^{+2.05}_{-1.99}$ \\[3pt]
$a_\mathrm{rel}$ (AU) & $1.00^{+0.04}_{-0.04}$ \\[3pt]
$M_2$ ($M_\mathrm{Jup}$) & $54.11^{+4.81}_{-4.36}$ \\[3pt]
$M_\mathrm{tot}$ ($M_\mathrm{Sun}$) & $0.104^{+0.013}_{-0.012}$ \\[3pt]
$M_1$ ($M_\mathrm{Sun}$) & $0.052^{+0.009}_{-0.008}$ \\[3pt]
$M_2$ ($M_\mathrm{Sun}$) & $0.052^{+0.005}_{-0.004}$ \\[3pt]
\hline
\multicolumn{2}{c}{Priors}\\
$\Delta\varpi$ (mas) & $0.34^{+0.07}_{-0.07}$ \\[3pt]
$\Delta I_\mathrm{FORS2}$ (mag) & $3.54^{+0.43}_{-0.40}$ \\[3pt]
\hline
\end{tabular}
\label{tab:solution}
\end{table}

\begin{table}
\caption{Results with $\Delta I_\mathrm{FORS2}'=\Delta I_\mathrm{FORS2}+1.5$ mag}
\centering
\begin{tabular}{cr}
\hline\hline
Parameter & Value \\
\hline
$\gamma$ (m s$^{-1}$) & $-10552.14^{+95.51}_{-100.11}$ \\[3pt]
\aphotocentre\ (mas) & $23.55^{+0.06}_{-0.06}$ \\[3pt]
$a_1$ (mas) & $24.31^{+0.41}_{-0.29}$ \\[3pt]
$a_\mathrm{rel}$ (mas) & $51.36^{+1.96}_{-1.90}$ \\[3pt]
$a_\mathrm{rel}$ (AU) & $1.00^{+0.04}_{-0.04}$ \\[3pt]
$M_\mathrm{tot}$ ($M_\mathrm{Sun}$) & $0.104^{+0.012}_{-0.011}$ \\[3pt]
$M_1$ ($M_\mathrm{Sun}$) & $0.055^{+0.008}_{-0.007}$ \\[3pt]
$M_2$ ($M_\mathrm{Sun}$) & $0.049^{+0.004}_{-0.004}$ \\[3pt]
\hline
$\Delta I_\mathrm{FORS2}'$ (mag) & $4.56^{+0.50}_{-0.47}$ \\[3pt]
\hline
\end{tabular}
\label{tab:solution2}
\end{table}

\begin{table}
\caption{Results with $\Delta I_\mathrm{FORS2}''=\Delta I_\mathrm{FORS2}+10$ mag}
\centering
\begin{tabular}{cr}
\hline\hline
Parameter & Value \\
\hline
$\gamma$ (m s$^{-1}$) & $-10447.03^{+87.16}_{-90.73}$ \\[3pt]
\aphotocentre\ (mas) & $23.54^{+0.06}_{-0.06}$ \\[3pt]
$a_1$ (mas) & $23.54^{+0.06}_{-0.06}$ \\[3pt]
$a_\mathrm{rel}$ (mas) & $51.17^{+1.85}_{-1.88}$ \\[3pt]
$a_\mathrm{rel}$ (AU) & $1.00^{+0.04}_{-0.04}$ \\[3pt]
$M_\mathrm{tot}$ ($M_\mathrm{Sun}$) & $0.103^{+0.011}_{-0.011}$ \\[3pt]
$M_1$ ($M_\mathrm{Sun}$) & $0.055^{+0.008}_{-0.007}$ \\[3pt]
$M_2$ ($M_\mathrm{Sun}$) & $0.047^{+0.003}_{-0.003}$ \\[3pt]
\hline
$\Delta I_\mathrm{FORS2}''$ (mag)& $12.81^{+0.53}_{-0.52}$ \\[3pt]
\hline
\end{tabular}
\label{tab:solution3}
\end{table}

\section{Discussion and interpretation}
\subsection{Absolute Magnitudes and Colour}
The 2MASS $K_S$-band magnitude of \dw\ is 11.46$\pm$0.03\,mag \citep{Cutri:2003nx}. Our absolute parallax implies a distance of $19.478\pm0.034$\,pc, for a system absolute magnitude of $M_{K_S}$ = 10.01$\pm$0.03\,mag. Applying our measured relative magnitude of 1.74$\pm$0.06\,mag in the Keck/NIRC2 $K_S$-band filter, we infer component absolute magnitudes $M_{K_S} = 10.17\pm0.03$\,mag for the primary and $M_{K_s} = 12.04\pm0.11$\,mag for the secondary.

The primary is thus pretty much spot on for an M8.5. The secondary is 1.7 mag fainter in $K$, thus faint enough that it could correspond to an L5-L6 dwarf. 

While the absolute brightness of \dw\ is consistent with expectations, the near-infrared colour and spectrum of this source are unusual. Its 2MASS $J-K_s = 1.22\pm0.16$ mag is relatively red compared to optically-classified M8 dwarfs (${\langle}J-K_s{\rangle} = 1.03 \pm 0.17$ mag; \citealt{2015AJ....149..158S}). This is clearly apparent in low-resolution near-infrared spectral data of \dw\ from \citet{2019ApJ...883..205B}, which has a consistent spectrophotometric color of $J-K_s$ = 1.26 and diverges
from the M8 dwarf standard VB~10 beyond 1.3~$\mu$m (Figure~\ref{fig:classify}). We found even worse agreement with intermediate-gravity (INT-G or $\beta$) and very low-gravity (VL-G or $\gamma$) M8 standards, consistent with the field gravity classification based on the methods of \citet{Allers:2013aa}. 

\subsection{Spectral binary fit}
\begin{figure}
\centering
\includegraphics[width=0.9\linewidth, trim=0 12mm 0 0, clip=true]{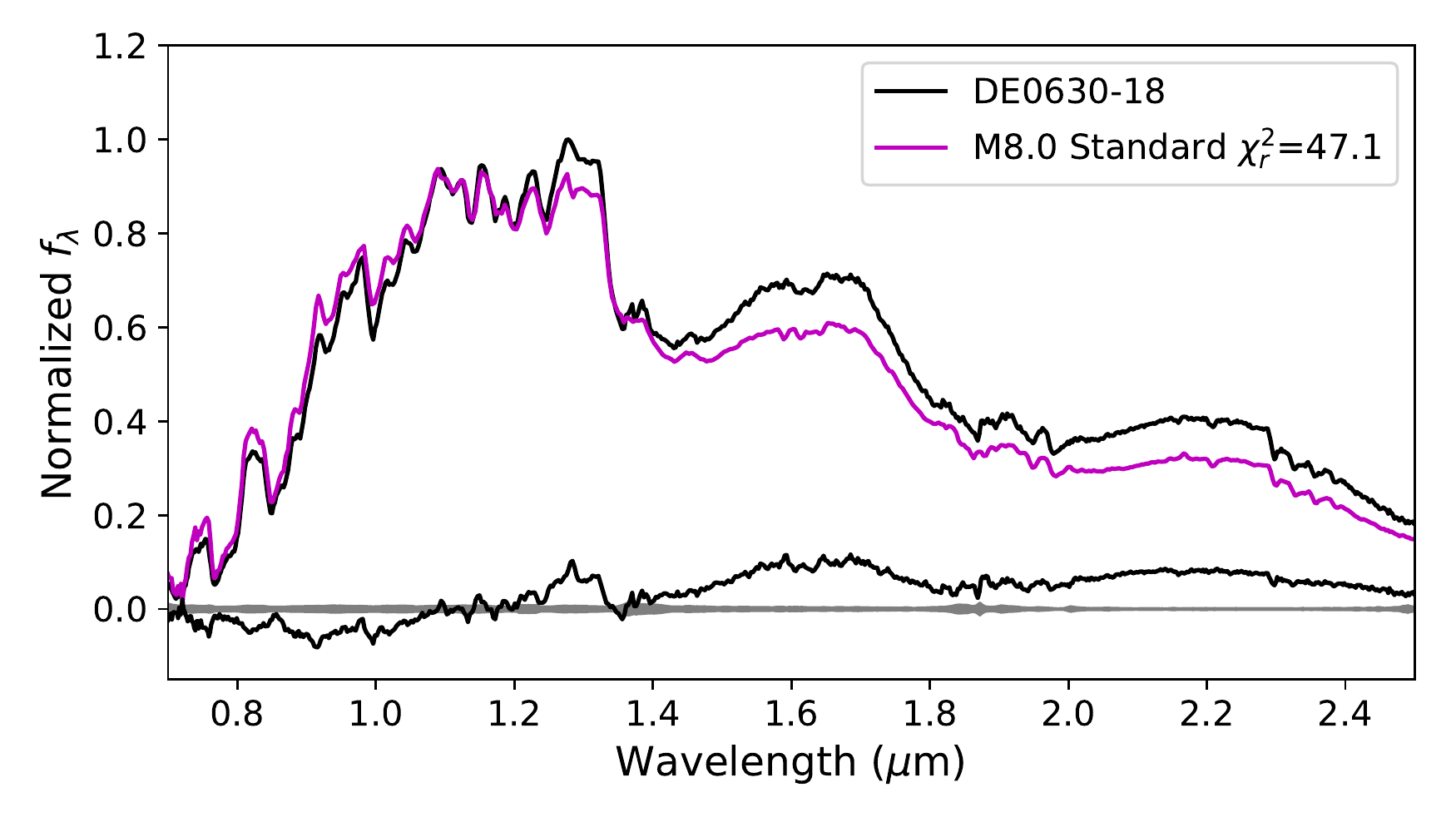}
\includegraphics[width=0.9\linewidth, trim=0 12mm 0 0, clip=true]{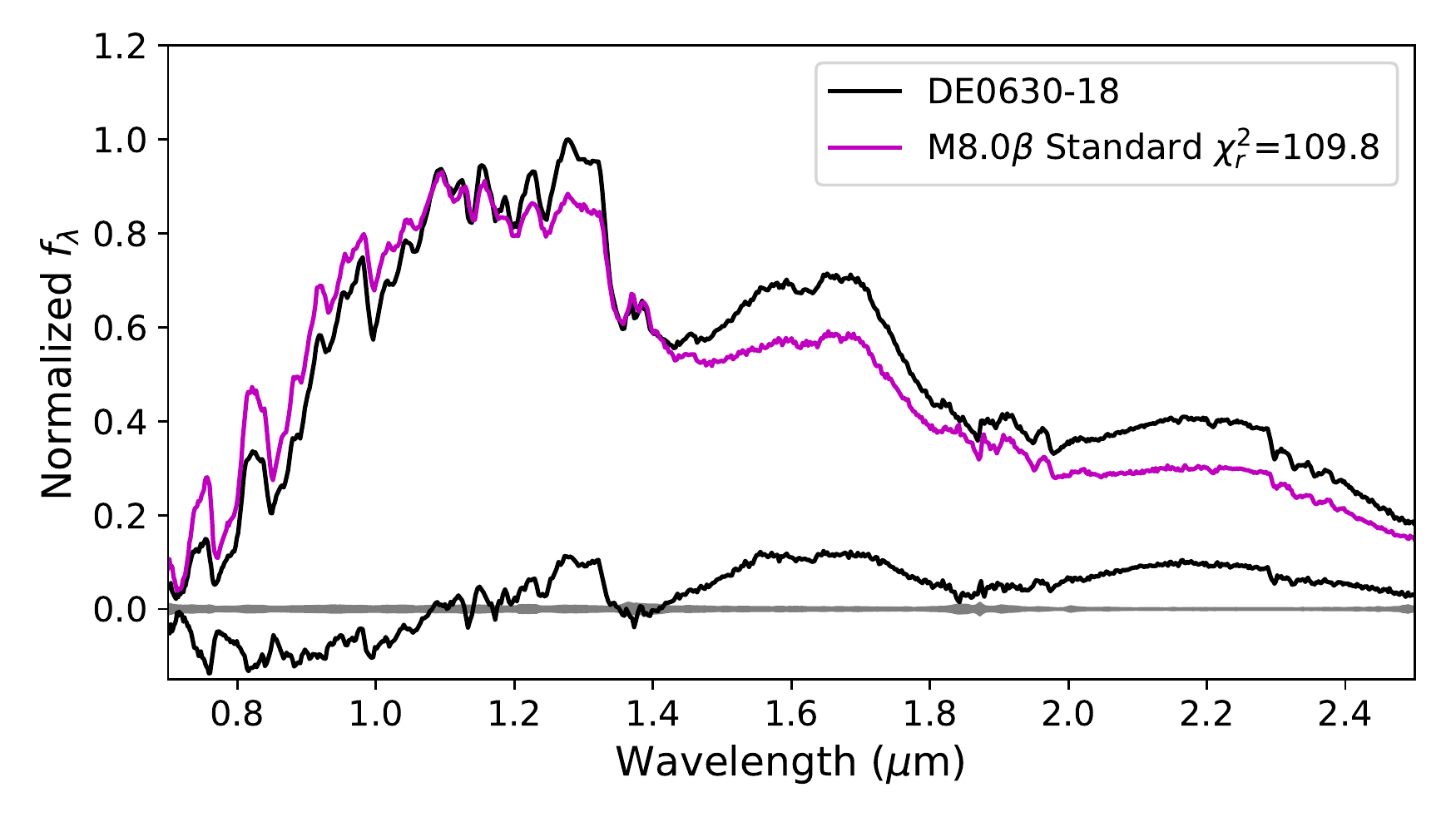} 
\includegraphics[width=0.9\linewidth, trim=0 12mm 0 0, clip=true]{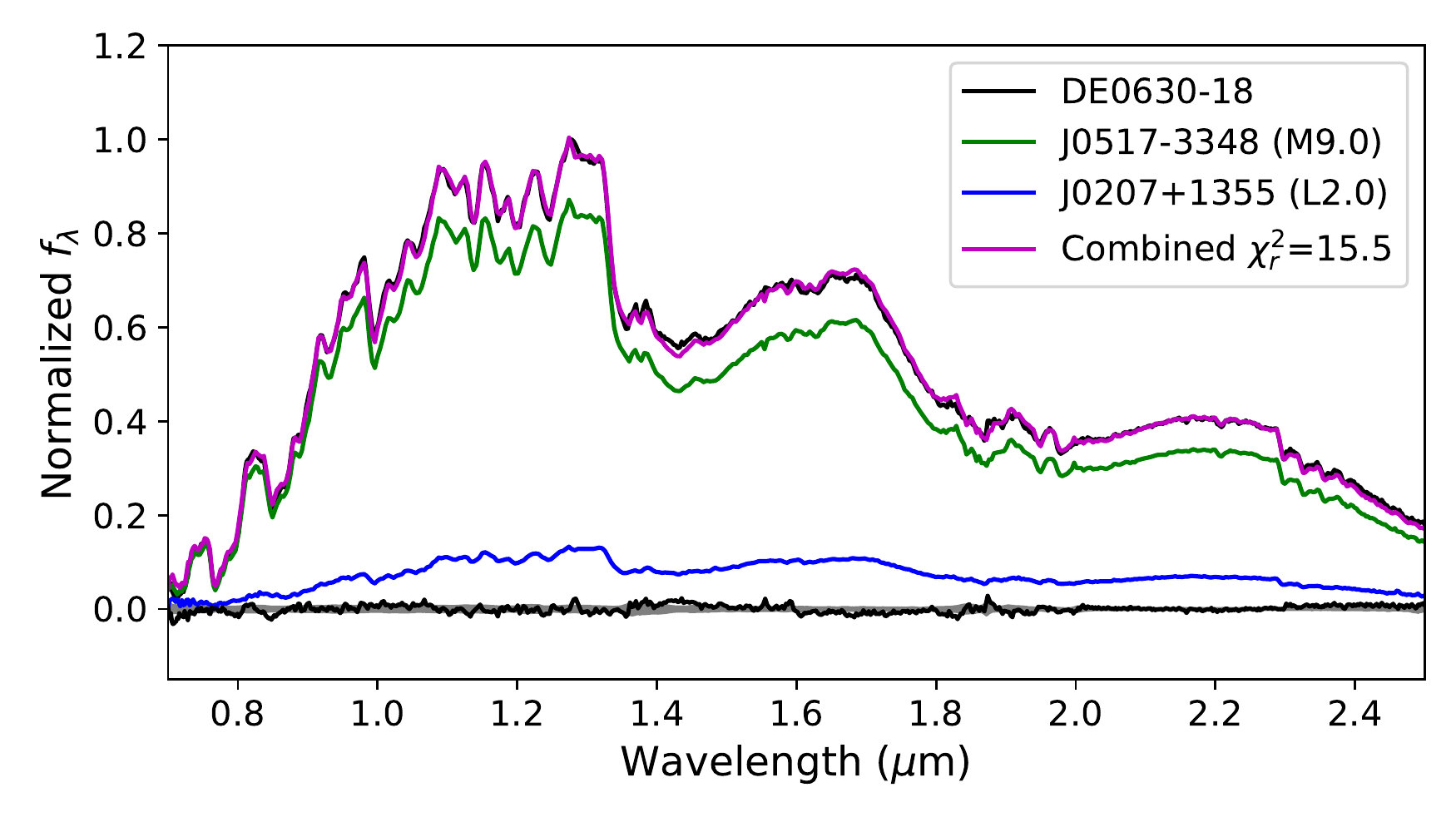} 
\includegraphics[width=0.9\linewidth, trim=0 12mm 0 0, clip=true]{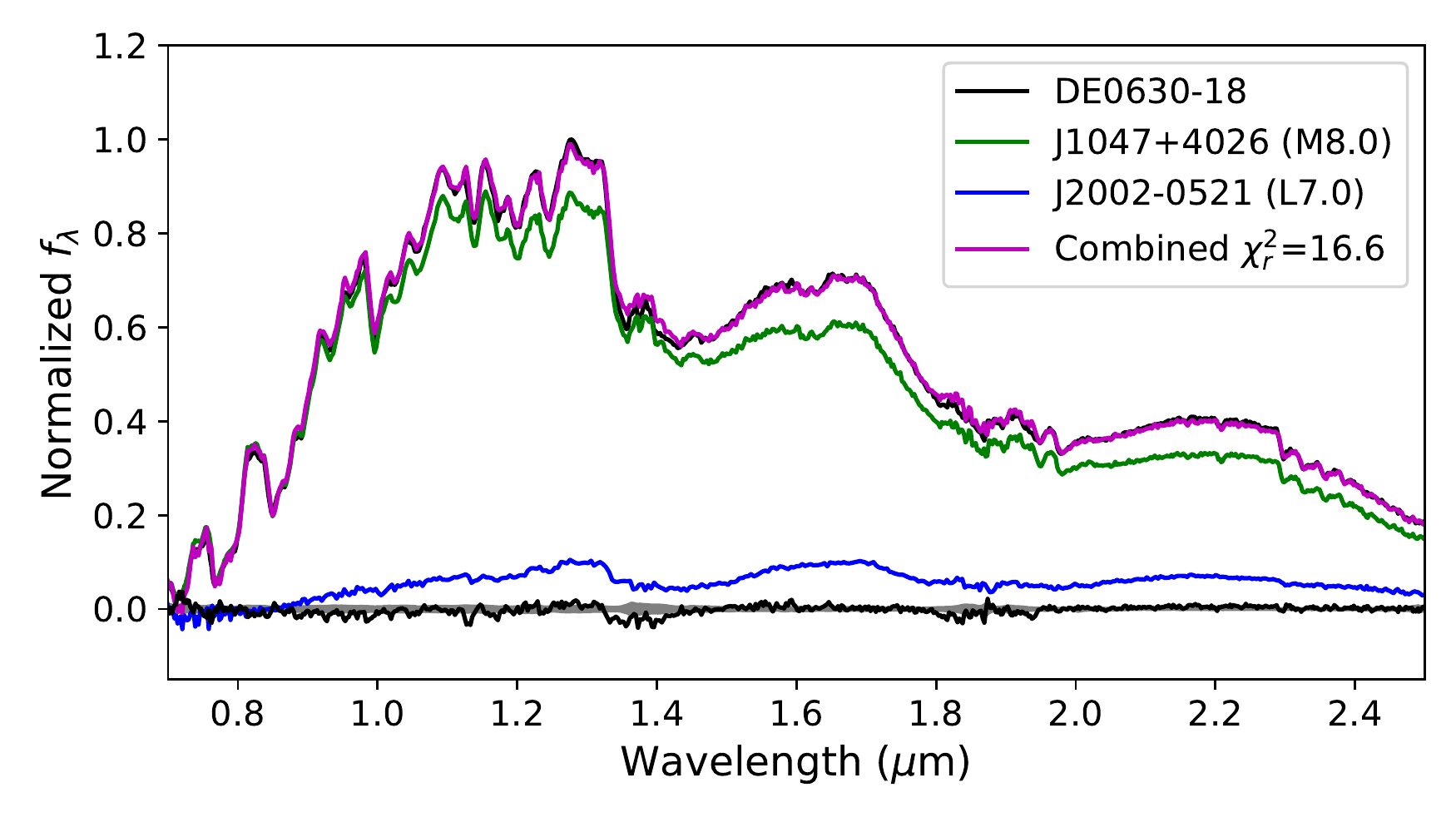} 
\includegraphics[width=0.9\linewidth]{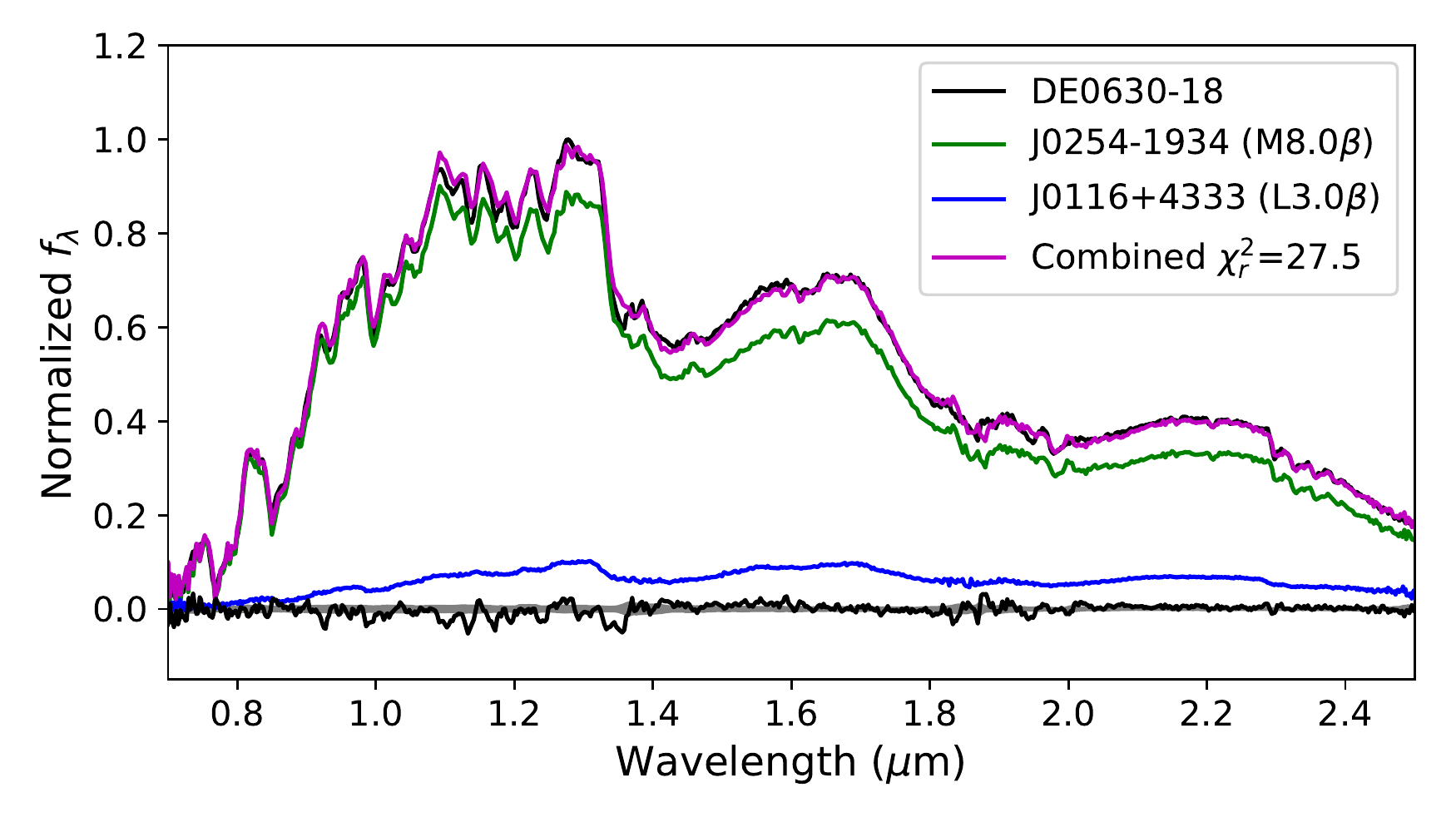} \\
\caption{From top to bottom, comparison of \dw's near-infrared spectrum (black lines) to various spectral templates (magenta lines): 
the M8 dwarf standard VB~10 (data from \citealt{Bardalez-Gagliuffi:2014aa_}), 
the intermediate-gravity M8$\beta$ standard 2MASSI~J0019262+461407 (data from \citealt{2019ApJ...883..205B}), 
the best-fit blended-light spectrum composed of a  field gravity late-M primary (green line) and early-L secondary (blue line), 
the best-fit blended-light spectrum composed of a  field gravity late-M primary (green line) and late-L secondary (blue line), 
and the best-fit blended-light spectrum composed of an intermediate-gravity late-M primary (green line) and L dwarf secondary (blue line).
The blended-light templates are all scaled to a relative magnitude of $\Delta{K_s}$ = 1.74.
All panels show the $\pm$3$\sigma$ uncertainties of the \dw\ spectrum (grey region) and difference spectrum (black line) at a median value of zero.}
\label{fig:classify}
\end{figure}

The most obvious explanation for this deviation is the contribution of secondary light to the blended-light spectrum. We compared the spectrum of \dw\ to three sets of binary templates constructed from data drawn from the SpeX Prism Library \citep{2014ASInC..11....7B}. 
We created two ``field'' template sets, using high signal-to-noise spectra (S/N $\geq$ 100) of 285 M7-M9 dwarfs for the primaries, and either S/N $\geq$ 75 spectra of 215 L1--L5 dwarfs (``early-type secondary'') or S/N $\geq$ 30 spectra of 79 L5-L9 dwarfs (``late-type secondary'') for the secondaries. We created a third ``young'' binary template set by combining S/N $\geq$ 70 spectra of 104 M7-M9 and S/N $\geq$ 30 spectra of 78 L3-L7 intermediate-gravity dwarfs. The gravity classifications of all templates were confirmed using the index-based scheme defined in \citet{Allers:2013aa}. The spectral components of all template binaries were scaled to match the observed $\Delta{K_s}$ = 1.74\,mag from the aperture masking observations. The bottom three panels in Figure~\ref{fig:classify} show that field-gravity and intermediate-gravity binary templates reproduce the spectrum of \dw\ significantly better than the single standards. The field template binaries provide better fits than the intermediate-gravity template, which shows low-level deviations across the 0.9--1.3~$\micron$ region, likely arising from mismatches in FeH band and atomic line strengths. {The field templates with early-type secondaries ($\sim$L2) and late-type secondaries ($\sim$L7) provide similar agreement, so these fits are insufficient to firmly determine the classification of the secondary, although the latter case would be consistent with an unresolved secondary component (see Section \ref{sec:interpretation}).} 

\subsection{Moving group membership}
Using our proper motion, parallax, and RV determinations of \dw\ we probed kinematic membership in known moving groups using the BANYAN tool\footnote{\url{http://www.exoplanetes.umontreal.ca/banyan}} \citep{2018ApJ...856...23G}, which yielded a 99.9 \% probability that \dw\ is a field object, thus there is no indication that it belongs to any of the considered young moving groups.  

\subsection{Lithium}
Despite the low masses in the system, we do not find signs of lithium absorption in our UVES spectrum \citep{Sahlmann:2015ac}. We could place an upper limit of $\sim$1\AA\ on the Li~I equivalent width (pEW) by visual inspection of simulations that consisted of injecting a spectral signature in the observed spectrum. 

At ages younger than 200 Myr, the lithium depletion boundary is located at $M_{K_s} < 10.0$ mag \citep{Dahm:2015} and consequently the lack of lithium detection in the primary indicates an age older than 200 Myr for the system.  
Should the age of the system be older than 200 Myr but younger than 700 Myr, the secondary could have preserved its initial lithium content. Considering for example the case that the secondary would have a similar LiI equivalent width as that of Hya12, an L4 member in the Hyades cluster with a pEW of 8.5 \AA\ and an age of 650 Myr \citep{Martin:2018aa}. If the primary contributes only flux, and no lithium absorption at all, the lithium feature from the secondary would be diluted by a factor of 15.8 considering a magnitude difference of 3 at 670.8 nm. The lithium feature in the secondary could then appear as a moderately weak absorption with pEW of about 0.54 \AA , which could not be detected in our UVES spectrum. 

We note that a diluted lithium feature from the secondary could 
be clearly detected in a reasonable exposure time of about 1 hour with a similar instrumental setup as that used for observing Hya12 with OSIRIS at the 10.4-meter Gran Telescopio de Canarias. It would be interesting to attempt such observation in the future in order to derive additional constraints on the age of the system. 
\dw\ may thereby become a member of the exclusive club of brown dwarf binaries for which dynamical masses and lithium depletion factors can be determined for each component, joining systems such as GJ 569B \citep{Zapatero-Osorio:2005aa}. 

\subsection{Interpretation}\label{sec:interpretation}
We have shown that no matter what the true $\Delta{I}$ is, the primary mass remains very low. Assuming the primary is single, its $M_{K_s}\sim 10.0$ mag implies $\log{L_\mathrm{bol}} \sim -3.3$ dex, and a mass of 0.055 $M_\mathrm{sun}$ implies an age of $\sim$150 Myr from the SM08 models. This is quite young for being FLD-G, but not unprecedented \citep{2016ApJ...821..120A}. 
The problem is the companion. If it is single, its inferred $M_{K_s} \sim 12$ mag implies $\log{L_\mathrm{bol}} \sim -4.1$ dex, and thus would have a mass of $\sim$0.030~$M_\mathrm{sun}$ according to the SM08 models. This is inconsistent with the measured component mass, as well as the combined mass of the system. $\sim$0.104~$M_\mathrm{sun}$. Moreover, an equal mass system is in tension with the distinct spectral morphologies inferred from the blended light infrared spectrum.

The empirical mass-$M_{K_s}$ relationship of \citet{2019ApJ...871...63M} for $M_{K_s}\sim 10.0$ mag predicts a primary mass of 0.086~$M_\mathrm{sun}$ ($\pm3$\%), which is also well above the astrometric mass measured here. Even if this higher mass is correct, the total mass of the system we require a secondary mass of only 0.030~$M_\mathrm{sun}$, again below its measured astrometric mass and consistent with a $\sim$150 Myr-old system based on evolutionary models. 

One scenario that could resolve these discrepancies is for \dw\ to be a young triple system with an unresolved binary secondary. For an age of $\sim$150 Myr, a primary of mass 0.055~$M_\mathrm{sun}$ and an unresolved secondary with components each of mass $\sim$0.025~$M_\mathrm{sun}$ would be consistent with all observations. These include the measured masses, the relative $K$-band magnitudes and component types (primary $T_{eff}$ $\approx$ 2400~K, secondary $T_{eff}$ $\approx$ 1400~K), the absence of strong intermediate-gravity features in infrared spectra, and the lack of Li~I in the combined-light optical spectra. 
Substellar triples are already been identified in both the field and young moving groups \citep{2005AJ....129..511B,2013ApJ...778...36R,2017ApJS..231...15D,2020NatAs.tmp...43T}, and \dw\ may represent such a system at an early stage of evolution. 

\section{Conclusions}
We presented individual mass determinations of the components in the \dw\ system on the basis of photocentre astrometry monitoring, one relative position determination, and a five radial velocity observations. The results indicate two components with nearly identical masses firmly in the substellar regime. This result does not depend on the precise value of the components' magnitude difference in the optical, which is the only quantity that we did not measure directly but that we had to determine empirically.

Given the object's M8.5 optical spectral type and absolute magnitude, but lack of Li~I absorption, this suggests that \dw\ is a relatively young system ($\sim$200 Myr). The measured relative magnitude and distinct spectral components inferred from analysis of its combined-light spectrum are at tension with the mass measurements, but can be resolved if the secondary of \dw\ is itself an unresolved binary.

Additional observations, e.g.\ a second relative separation measurement and spectroscopy targeting the diluted lithium signature from the secondary, would be beneficial in clarifying the properties of \dw.

\section*{Acknowledgements}
This research made use of the databases at the Centre de Donn\'ees astronomiques de Strasbourg (\url{http://cds.u-strasbg.fr});
NASA's Astrophysics Data System Service (\url{http://adsabs.harvard.edu/abstract\_service.html}), the paper repositories at arXiv; 
the SpeX Prism Libraries
(\url{http://www.browndwarfs.org/spexprism}); and  \textsc{ASTROPY}, a community-developed core Python package for Astronomy \citep{Astropy-Collaboration:2013aa}. 
The authors also made use of SCIPY \citep{Jones:2001aa}, NUMPY \citep{Oliphant2007}, IPYTHON \citep{Perez2007}, and MATPLOTLIB \citep{hunter2007}.
This work has made use of data from the
ESA space mission \emph{Gaia} (\url{http://www.cosmos.esa.int/gaia}),
processed by the \emph{Gaia} Data Processing and Analysis Consortium
(DPAC,
\url{http://www.cosmos.esa.int/web/gaia/dpac/consortium}). Funding for the DPAC has been provided by national institutions, in particular the
institutions participating in the \emph{Gaia} Multilateral Agreement.
This research was funded by the Ministerio de Economia y Competitividad and the Fondo Europeo de Desarrollo Regional (FEDER) under grant  AYA2015-69350-C3-1-P. 
AJB acknowledges funding support from the National Science Foundation under award No.\ AST-1517177. The material is based upon work supported by the National Aeronautics and Space Administration under Grant No.\ NNX15AI75G. M.\ C.\ L.\ acknowledges support from National Science Foundation grant AST-1518339.
The authors recognize and acknowledge the very significant cultural role and reverence that the summit of Mauna Kea has always had within the indigenous Hawaiian community. We are most fortunate and grateful to have the opportunity to conduct observations from this mountain.

\section*{Data availability}
The data underlying this article will be shared on reasonable request to the corresponding author.

\bibliographystyle{mnras}
\bibliography{biblio}

\bsp	
\label{lastpage}
\end{document}